\documentclass[sigconf,10pt]{acmart}
\AtBeginDocument{%
  \providecommand\BibTeX{{%
    \normalfont B\kern-0.5em{\scshape i\kern-0.25em b}\kern-0.8em\TeX}}}

\acmYear{2023}\copyrightyear{2023}
\acmConference[EuroMLSys '23]{3rd Workshop on Machine Learning and Systems}{May 8, 2023}{Rome, Italy}
\acmBooktitle{3rd Workshop on Machine Learning and Systems (EuroMLSys '23), May 8, 2023, Rome, Italy}
\acmPrice{15.00}
\acmDOI{10.1145/3578356.3592593}
\acmISBN{979-8-4007-0084-2/23/05}
\usepackage{caption}
\usepackage{subcaption}
\usepackage{algorithmicx}
\usepackage{algorithm}
\usepackage{algpseudocode}

\begin{document}

\title{Causal fault localisation in dataflow systems}

\author{Andrei Paleyes}
\email{ap2169@cam.ac.uk}
\affiliation{%
  \institution{Department of Computer Science and Technology, University of Cambridge}
  \city{Cambridge}
  \country{UK}
}

\author{Neil D. Lawrence}
\email{ndl21@cam.ac.uk}
\affiliation{%
  \institution{Department of Computer Science and Technology, University of Cambridge}
  \city{Cambridge}
  \country{UK}
}


\begin{abstract}
Dataflow computing was shown to bring significant benefits to multiple niches of systems engineering and has the potential to become a general-purpose paradigm of choice for data-driven application development. One of the characteristic features of dataflow computing is the natural access to the dataflow graph of the entire system. Recently it has been observed that these dataflow graphs can be treated as complete graphical causal models, opening opportunities to apply causal inference techniques to dataflow systems. In this demonstration paper we aim to provide the first practical validation of this idea with a particular focus on causal fault localisation. We provide multiple demonstrations of how causal inference can be used to detect software bugs and data shifts in multiple scenarios with three modern dataflow engines.
\end{abstract}



\maketitle

\section{Introduction}
Dataflow computing is an architectural paradigm for building software systems with a strong focus on data and its transformations. By expressing all logic of a system in terms of operations on data states and their data dependencies, developers following dataflow architecture produce a dataflow graph of an entire system as a natural part of the design process. The advantages of having direct access to the dataflow graph of a software application were already demonstrated in machine learning workloads \cite{barham2022pathways}, drug discovery \cite{lampa2016towards}, Internet of Things \cite{belsa2018flow}, programming education \cite{mason2017block}.

Recently it has been observed that there is a strong connection between dataflow graphs and structured causal models \cite{paleyes2023dataflow}. Importantly, it was noted that the dataflow systems produce graphs that are complete by design. This removes the need to build these graphs retroactively, relying on expert knowledge or discovery heuristics, thus resolving one of the biggest obstacles to the application of causal inference \cite{dawid2010beware}. The ability to treat graphs of dataflow systems as complete causal graphs has the potential to simplify applications of causal inference to dataflow systems. Paleyes et al. \cite{paleyes2023dataflow} hypothesised how causal inference can be used to aid technical troubleshooting, business analysis and experimentation in software systems that are designed with the dataflow paradigm. Nevertheless, the ideas proposed were not evaluated on real-life software, and thus require further validation. Our work continues this line of research, specifically focusing on using causal inference for fault localisation in dataflow systems. To that end, in this paper we present a demonstration study in which we aim to show how the causal attribution technique can be used with modern dataflow frameworks.

Our work is not the first to consider applications of causality to software systems. Causal testing tool Holmes \cite{Johnson2020CausalTU} generates test cases using input perturbations and then applies counterfactual causality to help developers identify sources of buggy behaviour in programs. Feature effects and interactions can be estimated with the notion of feature causality \cite{Dubslaff2022Causality}. BoGraph uses causal structure learning to auto-tune complex parametrised systems \cite{Alabed2022BoGraphSB}. Closer to the topic of our work, causality was used for fault localisation in software on method \cite{shu2013mfl, kuccuk2021improving} and program levels \cite{Baah2011MitigatingTC}, as well as in cloud computing \cite{aggarwal2021causal} and microservices \cite{ikram2022rca}. A recent review of applications of causality in software engineering observed named the discovery of the underlying dataflow graph as one of the key challenges  \cite{siebert2022applications}. Dataflow architecture resolves this problem, as it produces a dataflow graph as a part of the design process. This emphasises the need for a dedicated study of causal fault localisation for dataflow systems, and our work is the first step in that direction.

This paper presents a demonstration study in which we use the causal attribution mechanism proposed by Paleyes et al. \cite{paleyes2023dataflow} for fault localisation. By using the variety of dataflow frameworks, application domains and failure modes we aim to make the first step towards validating the feasibility of applying causal inference to dataflow systems.

\section{Causal inference on dataflow graphs}
Given that we provide a first real-life demonstration of concepts discussed by Paleyes et al. \cite{paleyes2023dataflow}, this section gives background information on the method proposed in that work.

Consider a dataflow system with a clearly defined output, and a distribution shift observed in the output. The task is to identify the most likely cause of the observed shift, which can be an issue in one of the system's components or a shift in one of the input data streams. The authors propose to approach this problem by modelling components of dataflow systems as causal conditionals. Given the observed shift in the output distribution, the idea is to recursively traverse the dataflow graph backwards, computing deviation (the node's output shift) and attribution score (how much the node contributes to the system's overall output shift) of each encountered data stream while modelling it as a random variable on which the downstream nodes are conditioned. Algorithm \ref{alg:attribution_change}, taken directly from the original work, implements this idea. We note that the fault in the system can occur due to either the change in an input data stream or in the computational logic, and the proposed technique can handle both cases.

\begin{algorithm}
\caption{Change attribution in software dataflow graphs, taken from Paleyes et al. \cite{paleyes2023dataflow}.}\label{alg:attribution_change}
\begin{algorithmic}[1]
\Require Dataflow graph with data streams $X_i$ and computational nodes $C_j$ ; $p \in \mathcal{D}^{new}, q \in \mathcal{D}^{old}$ distributions of all data streams in time frames $T^{new}, T^{old}$ respectively; target data stream $Y := X_d$
\State Model each data stream $X_i$ as a random variable.
\State Model each computational node $C_j$ with (input, output) data streams $(C_j^i, C_j^o)$ as conditional distributions.
\State Initialise a dictionary of attribution scores $F_{attr}$ with data stream nodes as keys.
\State Initialise an empty queue of data stream nodes $Q$.
\State Suppose we observe a change in $Y$ measured as $\Delta_Y = D(p(y)||q(y))$, where $D$ is KL divergence.
\State Push $Y$ to $Q$. 
\While{$Q$ is not empty}
\State Pop top element of $Q$ into a variable $S$
\State Find computational nodes $C_j$ that contribute to $S$
\State Compute deviation of $C_j$'s functionality: \newline
\hspace*{1.2em} $\Delta_{C_j} = \textbf{Deviation}(p(s|pa_s), q(s|pa_s))$
\For{each input data stream $ X_i \in C_j$}
\State Compute its contribution to the change in outcome: \newline
\hspace*{2.5em} $F_{attr}[X_i] = \textbf{Attribution}(X_i, \Delta_{C_j},  Y)$
\State Push $X_i$ to $Q$
\EndFor
\EndWhile
\State \Return $F_{attr}$

\end{algorithmic}
\end{algorithm}

The Algorithm \ref{alg:attribution_change} contains two subroutines: \textbf{Deviation} and \textbf{Attribution}. The purpose of the \textbf{Deviation} subroutine is to quantify the change of the output distribution in a given node. \textbf{Attribution} subroutine assigns each node a score that estimates the causal effect of a node on the shift of the system's output $Y$. Crucially these quantities should be computed in a way that allows for consistent comparison between nodes in the graph. Following Budhathoki et al. \cite{Budhathoki2021Why} and Singal et al. \cite{Singal2021Flow}, we use testing independence for \textbf{Deviation} subroutine and calculate Shapley values as scores in \textbf{Attribution}. We refer interested readers to \cite{paleyes2023dataflow} for a more in-depth discussion of the Algorithm \ref{alg:attribution_change}.

\section{Detailed example}
\label{section:detailed-example}
We begin demonstrating the potential utility of causal inference for fault localisation in dataflow programs by showing how it can be used to detect an unintentional change of behaviour in one of the components, also known as a software bug.

\begin{figure*}
	\centering
	\includegraphics[width=0.8\textwidth]{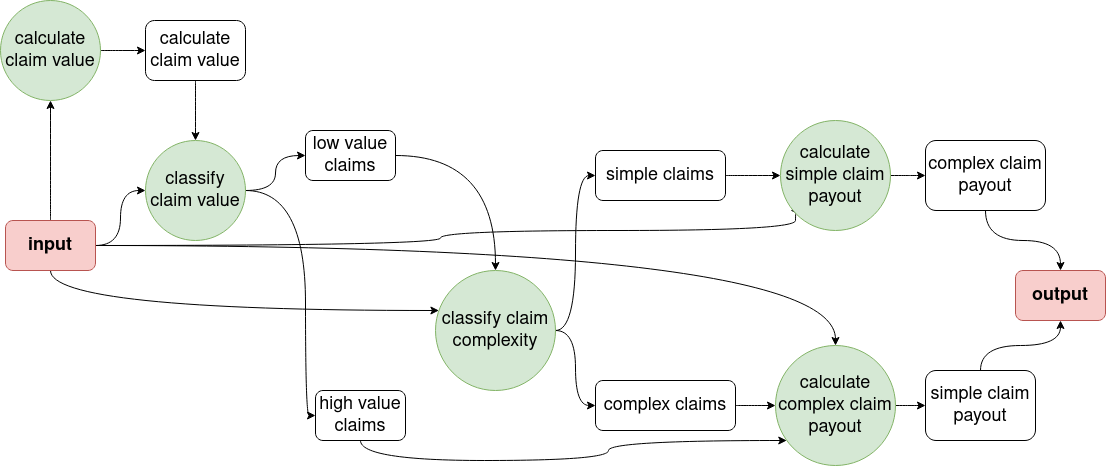}
	\caption{Dataflow graph of the Seldon Core v2 (SCv2) demo for processing car insurance claims. Underscores in component names are removed for readability. Green circles are processing nodes, red and white rectangles are data streams. Input and output streams are highlighted with red background and bold text.}
	\label{figure:scv2-graph}
\end{figure*}

The setting for this section is an application for processing insurance claims \cite{paleyes2022empirical}. The application processes car insurance claims by passing them through a series of data processing and decision making steps, finishing with calculating the payout for each claimant. The dataflow graph of our implementation of this application can be seen in Figure \ref{figure:scv2-graph}.

\begin{figure}
	\centering
	\includegraphics[width=\columnwidth]{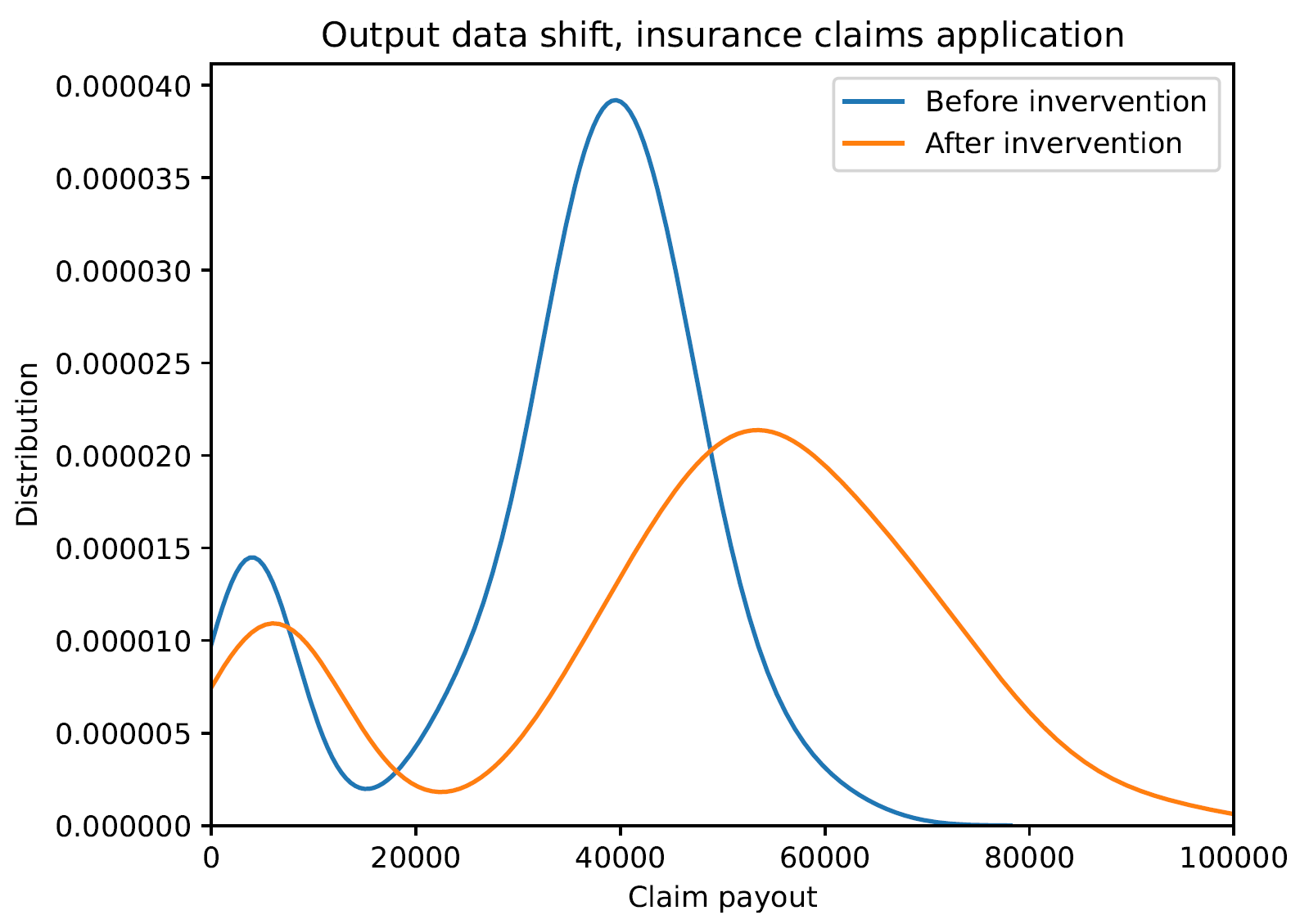}
	\caption{Payout amount distribution in the insurance claims application before and after an intervention. Payout is the final output of the claims processing pipeline, and detection of drift in its distribution shall trigger the fault localisation procedure.}
	\label{figure:insurance-claims-shift}
\end{figure}

As an intervention for this first experiment we create a bug in the system. Specifically, we change the behaviour of the component \textit{classify\_claim\_complexity} that classifies low value insurance claims as simple or complex. The bug causes this component to classify all claims as simple, thus impacting distributions of all downstream data nodes including the output stream. The shift of the output data distribution is shown in Figure \ref{figure:insurance-claims-shift}. We assume that the system has some kind of monitoring in place that detects the shift\footnote{Detection of data distribution shifts is a separate topic that we consider to be outside the scope of this paper. Interested readers can refer to Klaise et al. \cite{klaise2020monitoring} and  Quinonero-Candela et al. \cite{quinonero2008dataset}.}.

To identify the location of the bug, we build the dataset of data passing through the application by collecting raw data from each stream before and after the bug intervention and run the causal attribution procedure described in Algorithm \ref{alg:attribution_change}. We expect the \textit{classify\_claim\_complexity} component to receive the highest attribution score, identifying this node as the reason for the observed output distribution shift. Table \ref{table:scv2-bug-single-run} shows the result of applying the causal attribution procedure to identify the fault location. We display attribution scores for each data stream node of the application, and we also convert them to probabilities for convenience of comparison. 

First, we note that the causal attribution procedure correctly assigned the highest score to the \textit{simple\_claims} stream, which is an output stream of the \textit{classify\_claim\_complexity} node. A developer seeing this information can immediately deduce that since no other nodes contribute to the \textit{simple\_claims} data steam, it is its producing node that must be the problem. Moreover, we notice that since the attribution scores are assigned to every node, this method provides a quantification of uncertainty: the developer can estimate how confident the procedure is, and if any additional root cause investigation is required.

\begin{table}
\caption{Example of attribution scores and corresponding probabilities for a single run of fault localisation experiment. The highest score and probability are highlighted in bold font.}
\label{table:scv2-bug-single-run}
\begin{center}
\begin{tabular}{ |c|c|c|c|c| } 
    \hline
    Node Name & Score & Probability \\
    \hline
    input & 0.0014 & 0.043 \\
    calculate\_claim\_value & 0.0011 & 0.034 \\ 
    low\_value\_claims & 0.0011 & 0.034 \\ 
    high\_value\_claims & 0.0003 & 0.009 \\
    simple\_claims & \textbf{0.0163} & \textbf{0.503} \\ 
    complex\_claims & -0.0081 & 0.249 \\ 
    calculate\_simple\_claim\_payout & 0.0031 & 0.095 \\ 
    calculate\_complex\_claim\_payout & -0.0004 & 0.013 \\ 
    output & -0.0006 & 0.020 \\
    \hline
\end{tabular}
\end{center}
\end{table}

\section{Demonstrations}
In this section, we continue to demonstrate the causal fault localisation technique by showcasing experiments built with a variety of dataflow frameworks. We first discuss the frameworks themselves and then move on to the description of experiments. Code for all experiments can be found at \url{https://github.com/apaleyes/causality-fbp}, our implementation of the causal attribution procedure is done using DoWhy-GCM package \cite{blobaum2022dowhy}.

\subsection{Dataflow frameworks}
Here we give a brief overview of the dataflow frameworks used in our experiments, describing their general programming model and the features most relevant to our purposes. 

\textbf{Seldon Core v2 (SCv2)} is an open source framework for hosting machine learning models and running online and batch inference with them\footnote{\url{https://docs.seldon.io/projects/seldon-core/en/v2/index.html}}. Despite its strong focus on MLOps infrastructure, SCv2 can also be considered a general dataflow engine. SCv2 describes pipelines (also called ``inference graphs'') as a sequence of computations in Docker containers that communicate via Apache Kafka streams. Pipeline definitions are provided by the users in a form of YAML files with specifications of steps and connections. This computation model makes SCv2 decentralised and asynchronous by design \cite{akoush2022desiderata}, while also making it possible to access all intermediate data states of each inference request via Kafka streams. To facilitate debugging, traceability and data provenance SCv2 provides an ``inspect'' CLI utility that allows users to retrieve data for an entire pipeline as well as for individual steps. We make extensive use of the `inspect'' CLI to collect data for the experiments.

\textbf{Node-RED} is a flow-based visual programming tool particularly popular in the IoT space\footnote{https://nodered.org/}. Node-RED provides a lot of pre-made nodes that implement a large variety of functions, and users define the processing pipeline by placing nodes in the visual editor, specifying nodes' properties and connections between them. If a custom function is desired, it can be implemented as a custom node in JavaScript programming language. The ``msg Profiler'' node in the Node-RED ecosystem leverages the flow-based nature of the Node-RED programming model to provide the end-to-end tracing information for each message flowing through the system\footnote{\url{https://github.com/bartbutenaers/node-red-contrib-msg-profiler}}. We make use of this node to retrieve necessary operational information for each step of the dataflow graph.

\textbf{SciPipe} \cite{lampa2019scipipe} is a Go package for scientific workflows\footnote{\url{https://scipipe.org/}}. Workflows in SciPipe operate on files, with each workflow step being a command line expression that receives input file(s) and produces output file(s). The workflows themselves are described as Go programs. SciPipe's design closely follows principles of flow-based programming \cite{morrison1994flow}. SciPipe puts a strong emphasis on provenance and creates separate audit files for each successfully executed workflow step. Audit files contain metadata, such as timestamps, inputs and outputs, parent tasks. We make use of these audit files to discover connections between workflow steps, validate dataflow graph structure, and automate the analysis.

\begin{figure*}[t]
	\centering
	\includegraphics[width=1.01\textwidth]{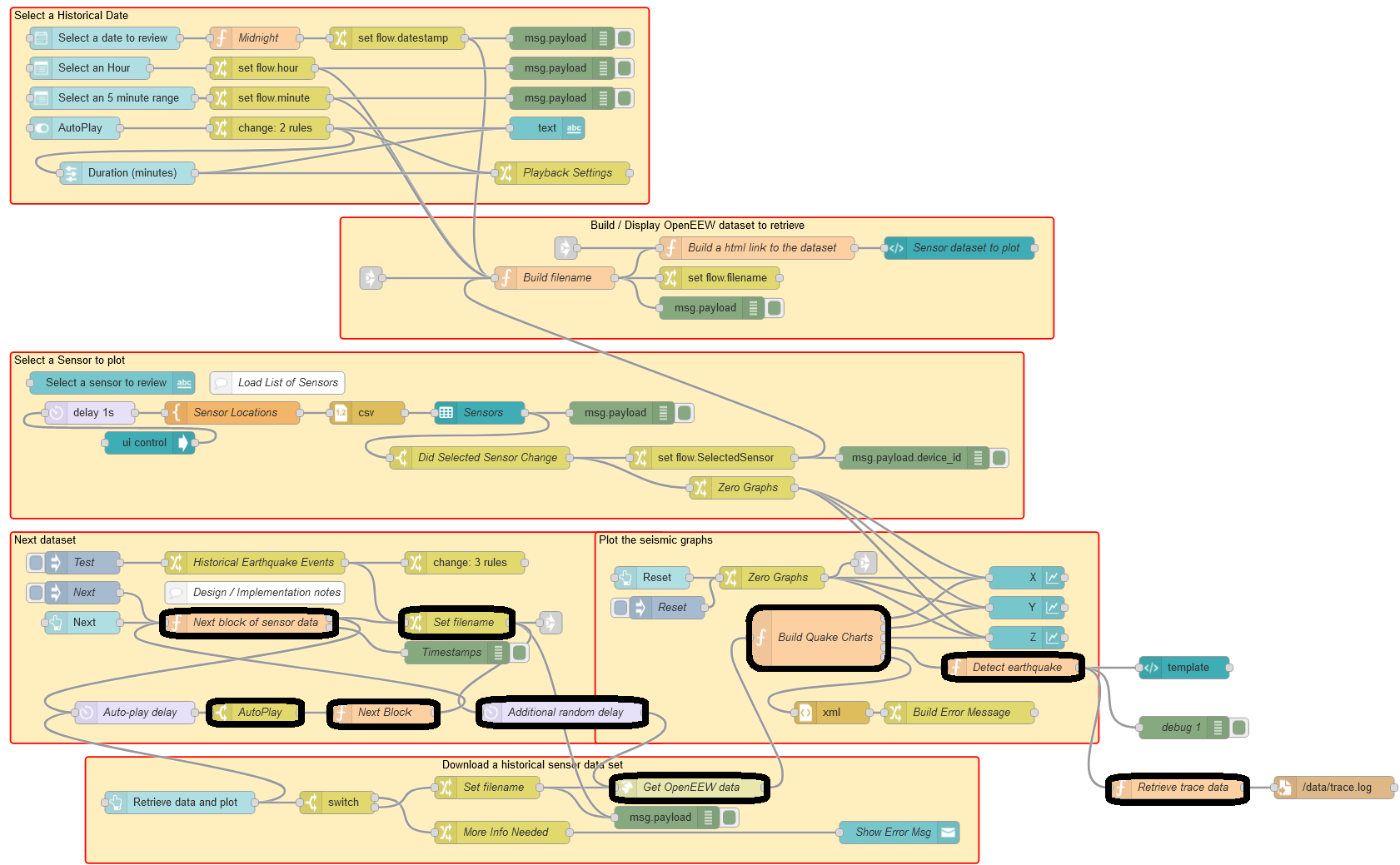}
	\caption{Dataflow graph of the Node-RED demo, as seen in the Node-RED editor. The demo is an interactive dashboard for monitoring earthquake data, majority of the nodes in the graph depicted here are responsible for UI elements and interaction with the user. The subgraph that is responsible for loading the data and analysing it is highlighted with a thick black border, and is the one we use in the experiment.}
	\label{figure:nodered-graph}
\end{figure*}

\subsection{Experiments}
\label{section:experiments}

We begin our evaluation of the causal fault localisation method with two experiments based on the insurance claims application described in the previous section, implemented with SCv2 and displayed in Figure \ref{figure:scv2-graph}. The first experiment involves the intervention in a form of a software bug and is already described in Section \ref{section:detailed-example}. In the second experiment, we intervene on the input data distribution, increasing the originally claimed amount by 50\%. We expect the attribution procedure to ignore distribution changes in all intermediate nodes and assign the highest score to the \textit{input} stream.

Our third experiment demonstrates that the causal fault localisation procedure can be used not only for business data but also for technical operational metrics. For that experiment we use the seismic activity monitoring dashboard built with Node-RED on OpenEEW data\footnote{\url{https://github.com/openeew/openeew-nodered}}. The dashboard allows a user to select time and location and then loads sensor data to examine for possible earthquake signals. Dashboard's dataflow graph as seen in the Node-RED interface is shown in Figure \ref{figure:nodered-graph}. The dashboard has an "auto-load" feature that continuously loads new data, and allows us to measure the throughput of the pipeline. We define the throughput of the system (node) as the number of messages that pass through the system (node) in a fixed time period of 5 seconds. We then insert a random sleep delay between 0.5s and 1s in the \textit{Build Quake Charts} node, observe the drop in the end-to-end throughput and expect the causal attribution procedure to correctly identify the node responsible for the drop in performance.

\begin{figure}
	\centering
	\includegraphics[width=\columnwidth]{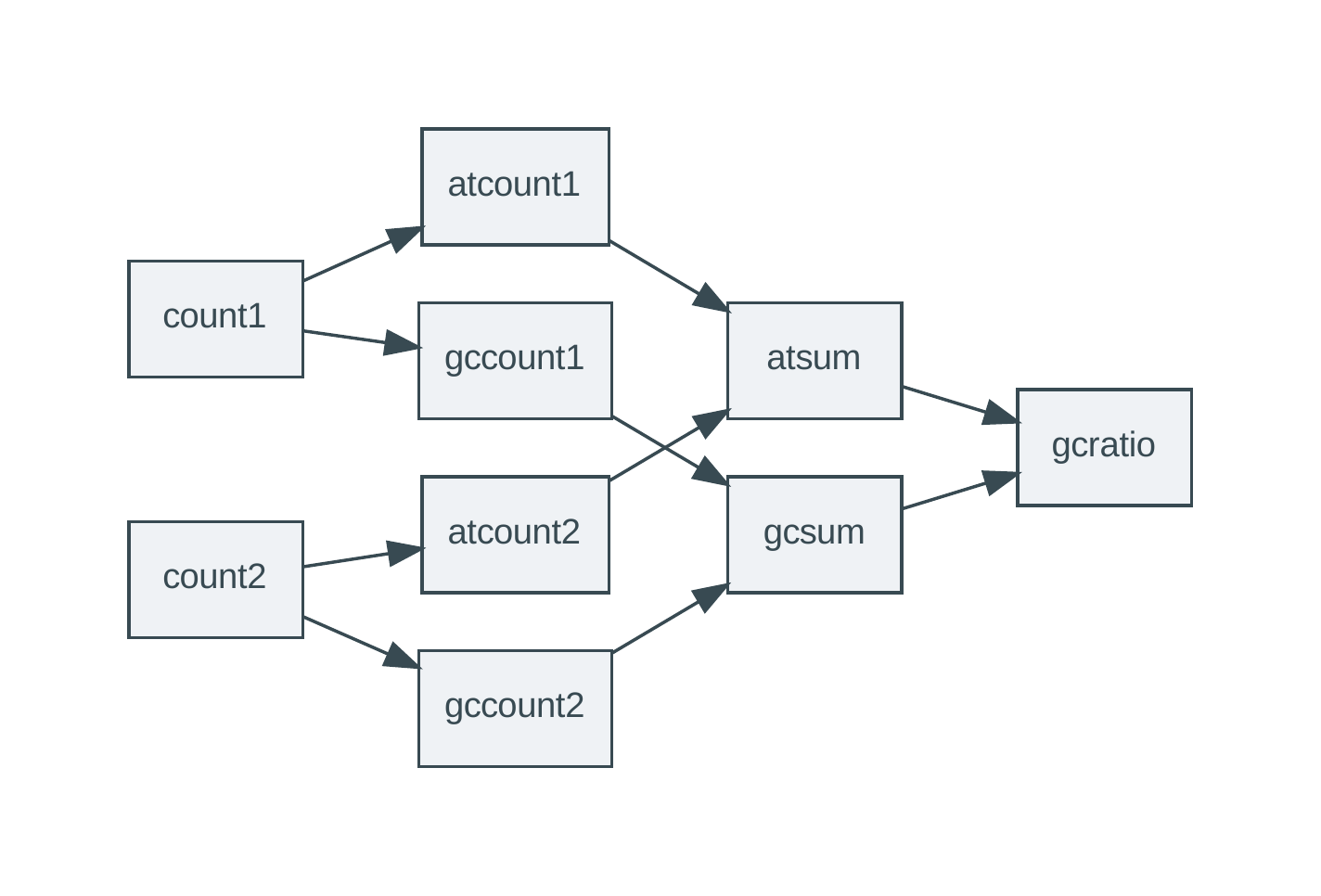}
	\caption{Dataflow graph of the SciPipe demo for calculating GC ratio across two segments of a DNA sequence. This graph's image is generated with the SciPipe API.}
	\label{figure:scipipe-graph}
\end{figure}

Our final experiments are designed to show that the causal fault localisation procedure can correctly identify drifting data stream among multiple input sources. For that, we are using the workflow we have implemented with SciPipe that computes the GC ratio across two segments of a DNA sequence. Two inputs are given to the workflow: the number of lines to collect from the first and second segments referred to as \textit{count1} and \textit{count2} respectively. The workflow then proceeds to extract the specified number of lines from each of the two segments, calculate AT and GC content in each, aggregate them and produce the final GC ratio. The dataflow graph of the workflow can be seen in Figure \ref{figure:scipipe-graph}.

To induce output drift in the workflow, in this example we intervene on the distribution of one of the input nodes. We expect our attribution procedure to identify the affected input source, ignoring the other input source and workflow steps. Since there are two different input sources in this workflow, named \textit{count1} and \textit{count2}, we run two separate experiments for this demo.

\begin{figure*}
	\begin{subfigure}[b]{1.03\columnwidth}
		\includegraphics[width=\columnwidth]{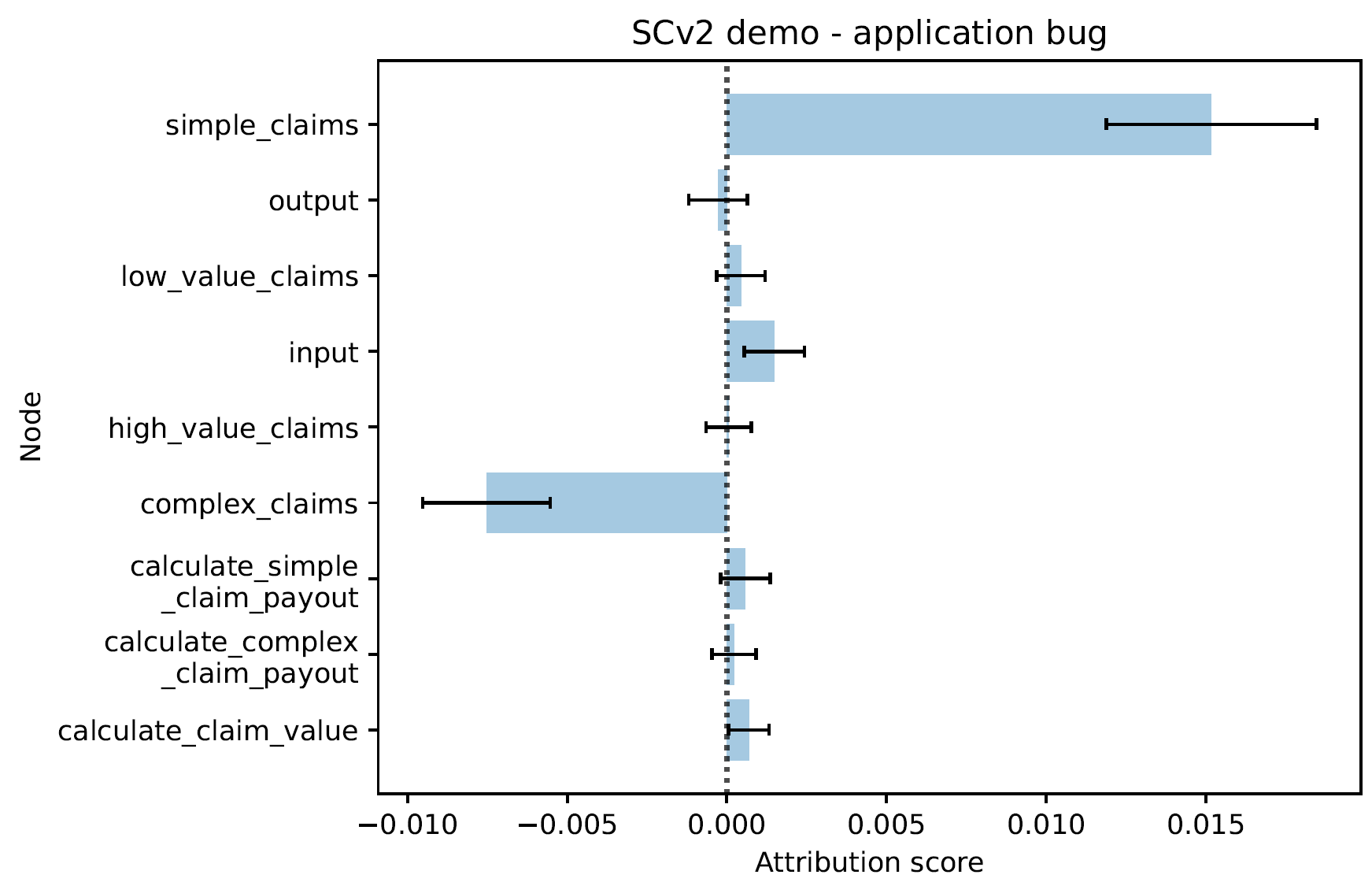}
		\caption{}
		\label{figure:scv2-bug}
	\end{subfigure}
	\begin{subfigure}[b]{1.03\columnwidth}
		\includegraphics[width=\columnwidth]{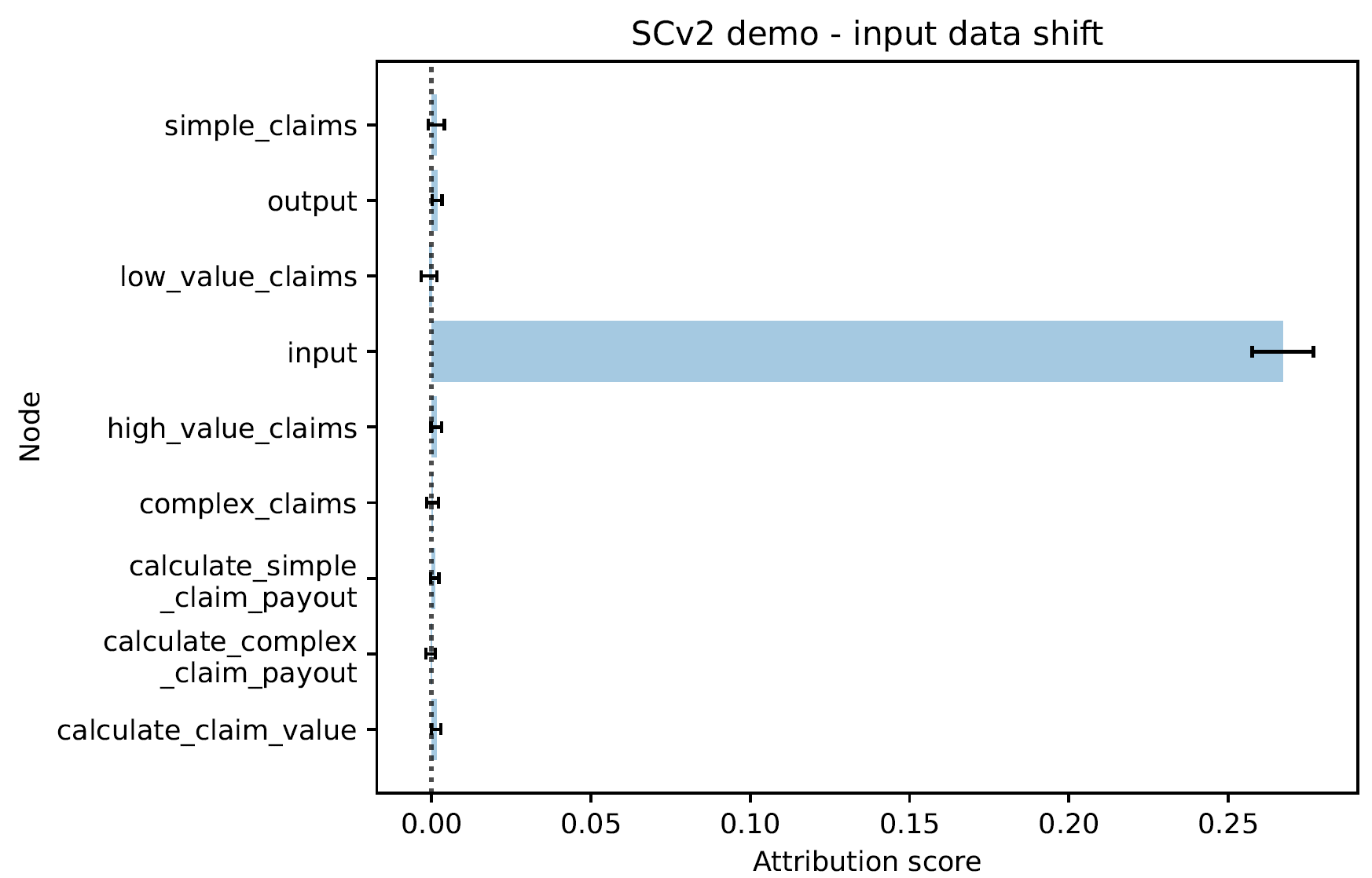}
		\caption{ }
		\label{figure:scv2-input}
	\end{subfigure}
	\begin{subfigure}[b]{1.00\columnwidth}
		\includegraphics[width=\columnwidth]{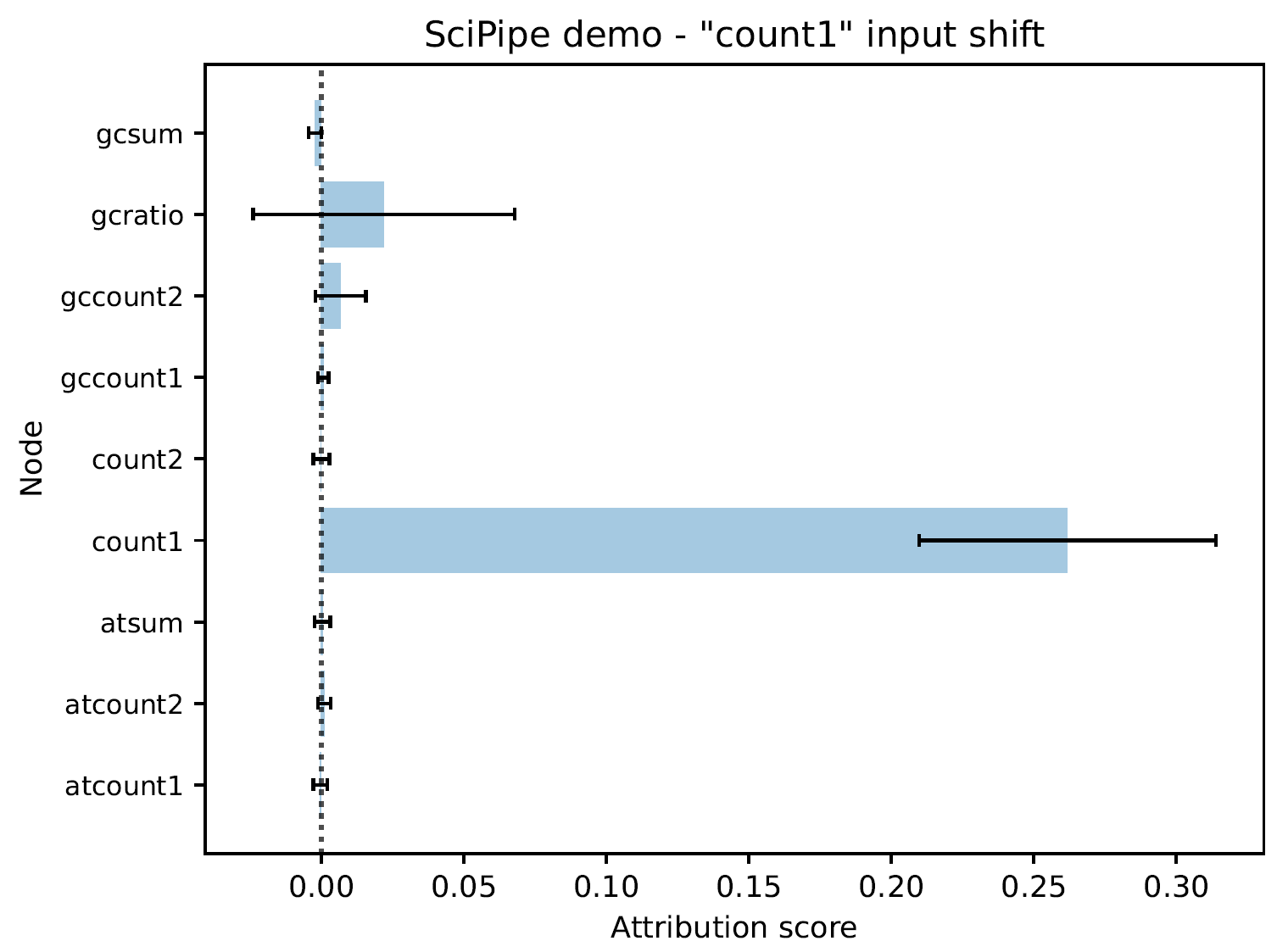}
		\caption{}
		\label{figure:scipipe-count1}
	\end{subfigure}
	\begin{subfigure}[b]{1.00\columnwidth}
		\includegraphics[width=\columnwidth]{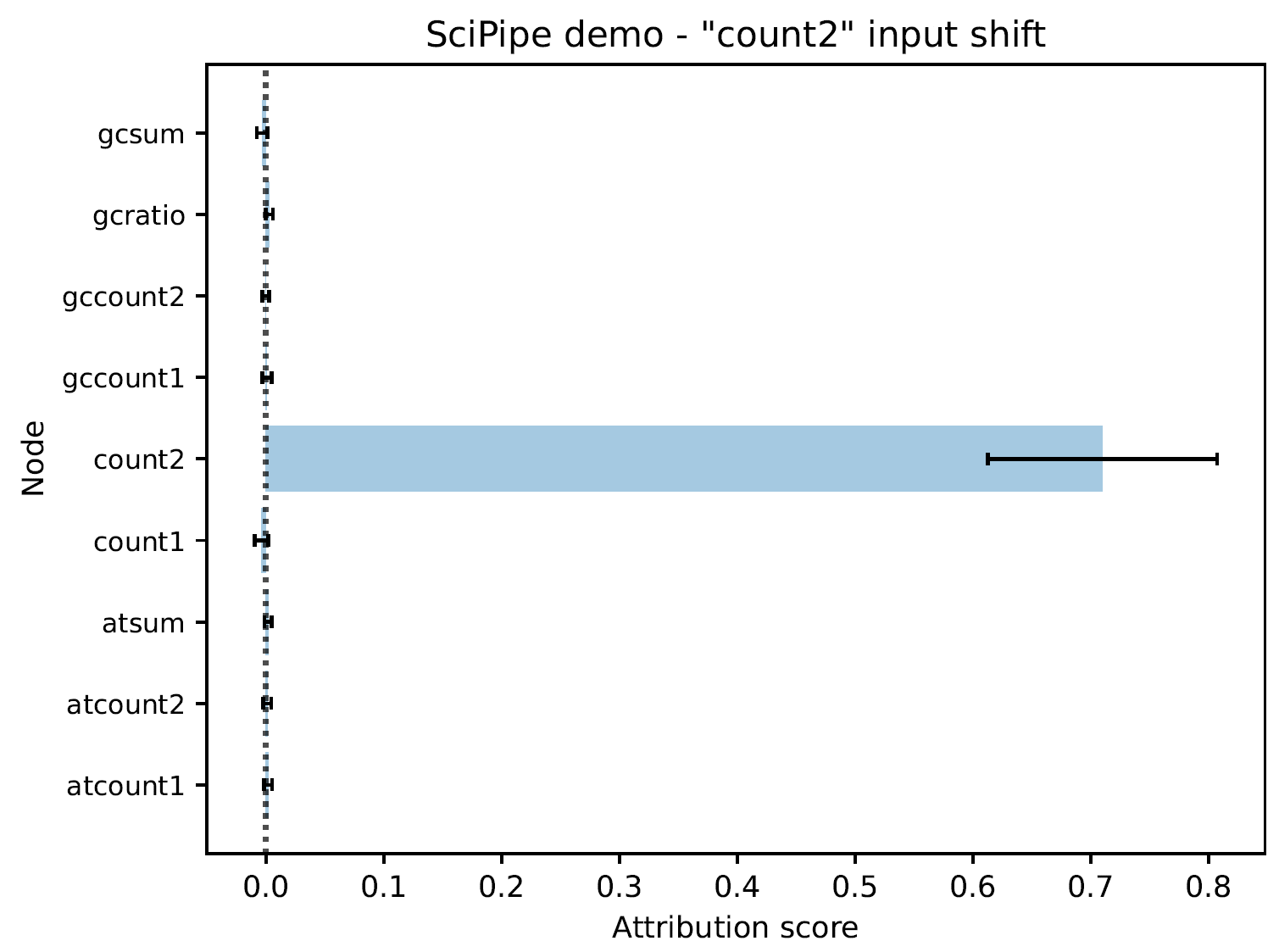}
		\caption{}
		\label{figure:scipipe-count2}
	\end{subfigure}
	\begin{subfigure}[b]{1.2\columnwidth}
		\includegraphics[width=\columnwidth]{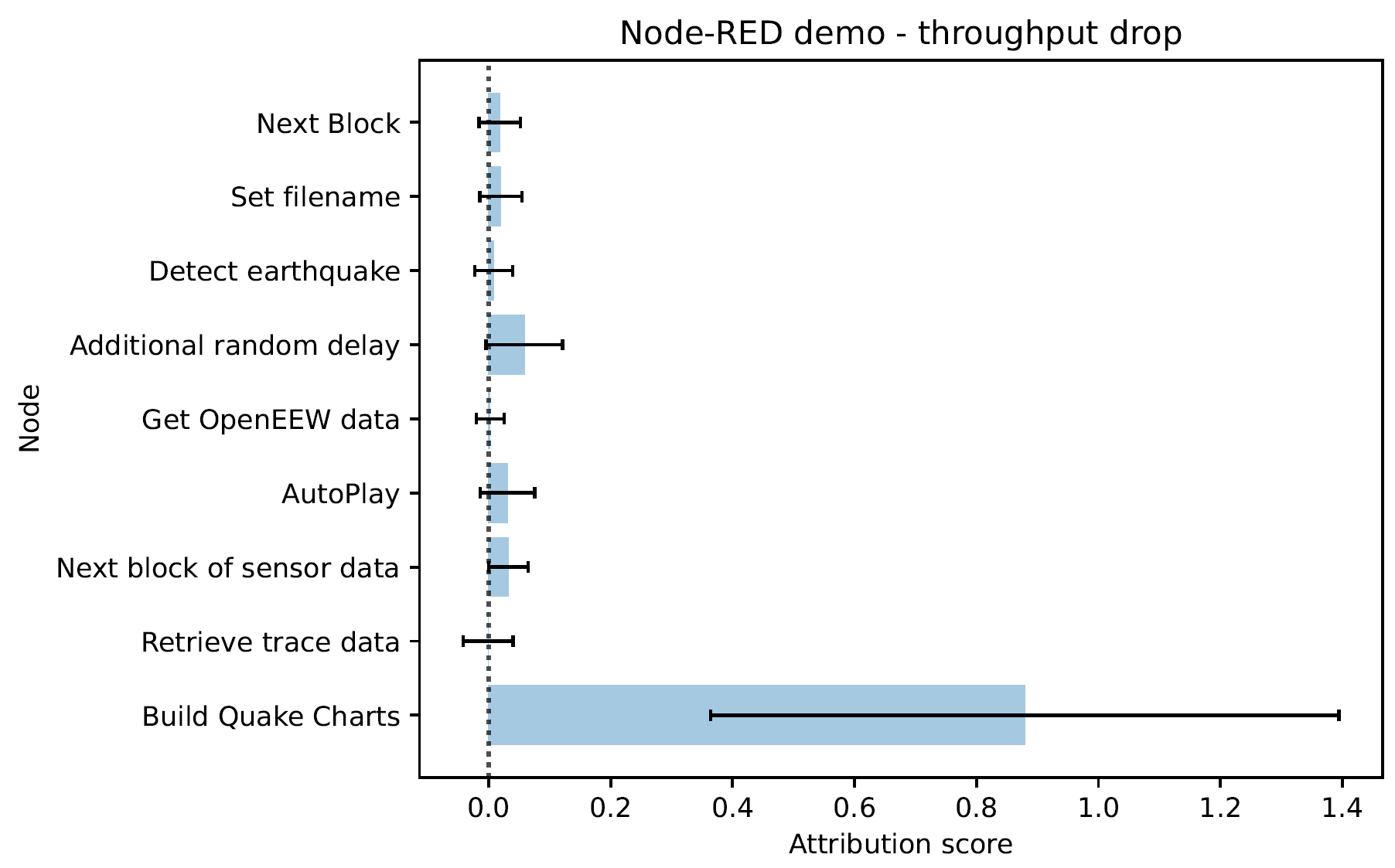}
		\caption{ }
		\label{figure:node-red-throughput}
	\end{subfigure}
	
	\caption{Attribution score distributions for all experiments across 20 repeats. For each node we report the mean attribution score and confidence intervals. In all cases the intervention node is correctly assigned the highest mean score. For all experiments we report $p<0.01$ on Welch's t-test for picking the correct offending node over any other node, supporting the statistical significance of our results.}
	\label{figure:attribution-scores}
\end{figure*}

Figure \ref{figure:attribution-scores} displays the results of running the described experiments. Each experiment is repeated 20 times on randomised input to verify its robustness, and the figures report mean values and confidence intervals of attribution scores for each node in the dataflow graph of an application. In all five experiments the causal attribution approach was able to correctly identify the offending component of the corresponding dataflow program. We have also used Welch's t-test to verify that our results are statistically significant and observed $p < 0.01$ in all experiments.

\section{Future work}
Although in all our experiments causal inference was able to correctly localise the faulty node, we think this technique is not ready yet for adoption in production-ready applications. This paper is merely the very first step towards the general availability of the causal fault localisation technique in dataflow systems. For instance, more efforts are likely required to validate its scalability. Graphs in our experiments do not exceed 20 nodes, while modern software can yield complex and large graphs with hundreds or even thousands of nodes \cite{musco2017large}. Similarly, the method we use assumes that the dataflow graph is acyclic, while it is normal for complex software workflows to include loops. Thus an extension of Algorithm \ref{alg:attribution_change} for graphs with cycles is desirable. Further, it is necessary to study cases where data flowing through the graph represents not a single record but a dataset, a use case commonly arising in machine learning pipelines.

Multiple immediate next steps can be made following our paper. One of the demonstrations described in Section \ref{section:experiments} can be turned into an automated utility for fault localisation in the respective dataflow framework. Since all frameworks we used are fully open-sourced, said utility could become a part of the framework's monitoring capabilities. Additionally, the original paper \cite{paleyes2023dataflow} postulated that causal inference can be used for business analysis and experimentation in dataflow systems (in addition to fault localisation we considered), and similar demonstrations are necessary to validate the viability of these claims.

In this paper, we study the use of causal inference techniques in dataflow systems. We believe dataflow architecture provides developers with multiple useful properties \cite{paleyes2021towards}, in addition to the applicability of causal inference, and encourage wider usage of this paradigm. Therefore our results are most relevant to systems built with dataflow frameworks, similar to Seldon Core, SciPipe or Node-RED. Our method relies on the fact that dataflow frameworks we used provide developers with complete and explicit dataflow graphs of an entire system. However, it might be possible to apply the same techniques in cases where only approximate graphs can be recovered, for example where existing software not built with a dataflow approach is being analysed. A similar demonstration study on such software systems might be of interest to the community.

\section{Conclusions}
In this demonstration paper, we showed how causal inference can be used for fault localisation in dataflow systems in a straightforward manner, thus continuing the line of research started by Paleyes et al. \cite{paleyes2023dataflow}. We have implemented five demonstrations exploring various possible failure modes in dataflow systems and showed how the same technique can be used across different dataflow engines. The source code of all experiments is openly available, enabling researchers to replicate and extend our work.

We hope this paper emphasises the synergy between causal inference and dataflow architecture, as well as makes a step towards the general applicability of causal fault localisation in dataflow software systems.


\bibliographystyle{ACM-Reference-Format}
\bibliography{references}

\end{document}